\documentstyle[eqsecnum,epsf,aps,twocolumn]{revtex}

\begin{document}
\draft
\preprint{HEP/123-qed}

%%%%%%%%%%%%%%%%%%%%%%%%%%%%%%%%%%%%%%%%%%%%%%%%%%%%%
\twocolumn[\hsize\textwidth\columnwidth\hsize\csname
@twocolumnfalse\endcsname
%%%%%%%%%%%%%%%%%%%%%%%%%%%%%%%%%%%%%%%%%%%%%%%%%%%%%

\title{ Anisotropy of magnetothermal conductivity in Sr$_2$RuO$_4$ }

\author{ M. A. Tanatar$^{1,2,}$\cite{email,ikhp}, M. Suzuki$^1$, S. Nagai$^1$, Z. Q. Mao$^{1,2}$, Y. Maeno$^{1,2}$, and T. Ishiguro$^{1,2}$}
\address{
$^1$ Department of Physics, Kyoto University, Kyoto 606-8502, Japan\\
$^2$ CREST, Japan Science and Technology Corporation, Kawaguchi, Saitama 332-0012, Japan}

\date{\today}
\maketitle
\begin{abstract}
The dependence of in-plane and interplane thermal conductivities of Sr$_2$RuO$_4$ on temperature, as well as magnetic field strength and orientation, is reported. We found no notable anisotropy in the thermal conductivity for the magnetic field rotation parallel to the conducting plane in the whole range of experimental temperatures and fields, except in the vicinity of the upper critical field $H_{\rm c2}$, where the anisotropy of the $H_{\rm c2}$ itself plays a dominant role. This finding imposes strong constraints on the possible models of superconductivity in Sr$_2$RuO$_4$ and supports the existence of a superconducting gap with a line of nodes running orthogonal to the Fermi surface cylinder.

\end{abstract}

]\narrowtext

The nature of superconductivity of Sr$_2$RuO$_4$ has attracted a lot of interest recently \cite{nature1,review}. An early theoretical prediction \cite{rice} about the spin-triplet character of the superconductivity of the material was supported by several key experimental findings made within the last few years. These include ($i$) the strong dependence of the superconducting transition temperature $T_{\rm c}$ on the density of non-magnetic impurities \cite{andyimp,defect}, ($ii$) the spontaneous appearance of magnetic field below $T_{\rm c}$ \cite{musr}, as expected for a superconductor with broken time reversal symmetry (TRS), and ($iii$) the lack of any change of the spin susceptibility in the magnetic field parallel to the conducting plane \cite{nmrnature,neutron}, as expected for a spin-triplet superconductor with the Cooper-pair spins parallel to the conducting planes. 

To explain experiments ($i$)-($iii$), a $p$-wave superconducting state with broken TRS, described by a {\bf $d$}-vector {\bf $d$}=$\hat{z}(k_x+ik_y)$, was proposed \cite{sigrist3}. This state has a full gap on the Fermi surface (FS), with the gap magnitude possibly being different on different FS sheets \cite{sigrist2}. However, recent experiments \cite{zakinew,nqr,penetration}, performed on very pure samples (characterized by $T_{\rm c}$ close to the estimated value for perfect material $T_{\rm c0}$ =1.5 K) revealed a tendency for a power-law behavior of quasi-particle excitations down to the lowest temperatures (50 mK), contrary to the exponential decrease for the fully gapped state. In view of these findings it was suggested that $d$- or $f$-wave states may be better choices for the description of the superconducting state in Sr$_2$RuO$_4$ \cite{hasegawa,balatsky,makianiz,mazin}, although the spin-singlet $d$- wave state would be incompatible with the experimental result ($iii$). 

For a correct description of the superconducting order parameter in Sr$_2$RuO$_4$, it is important to know the location of the line nodes, if there are any, on the FS. This information can be obtained only with the use of a directional probe. Study of the anisotropy of magnetic field dependence of thermal conductivity is one of the key experiments for this purpose. Similar experiments were performed on the high-$T_{\rm c}$ cuprates, giving a four-fold anisotropy consistent with the $d$-wave order parameter \cite{yuaniz,4foldcuprates}. 

In this Letter we report the study of thermal conductivity $\kappa$ of Sr$_2$RuO$_4$ as a function of the orientation of the magnetic field within the conducting plane. We studied the thermal conductivity with a heat flow direction, $\bar{Q}$, both along the plane, in [100] direction ($\kappa_{100}$), and perpendicular to the conducting plane ($\kappa_{001}$). $\kappa_{100}$ is electronic in origin \cite{Suderow,tanatar} and thus is directly related to the structure of the superconducting gap; however, among several factors limiting the accuracy of $\kappa_{100}$ anisotropy measurement, the lack of the symmetry in the magnetic field $H$ direction with respect to $\bar{Q}$ is quite inevitable. This makes inter-plane configuration preferable, since here $H$ parallel to the planes is always transverse to $\bar{Q}$. Study of phonon conductivity, which dominates in this configuration, offers additionally essential simplification of the understanding of quasiparticle density variation. In electronic thermal conductivity, generation of quasiparticles due to the Volovik effect \cite{Volovik} should be taken in a self-consistent way: on increasing $H$ the increase of the density of heat carriers and the decrease of their mean free path nearly cancel each other \cite{Kubert,Vekhter}. However, the density of heat carriers in phonon transport is determined only by temperature $T$ and is constant in $H$, hence all the variation with $H$ comes essentially from the increase in the density of scatterers, mainly quasiparticles in the bulk \cite{Suzuki}.

The main result of our study is that we did not find any anisotropy beyond the accuracy of our experiment, of the order of 0.5\%.This is valid in the whole experimental range of temperatures $T$ (down to 0.3 K) and magnetic fields, except very near to the upper critical field $H_{\rm c2}$. A two-fold anisotropy up to 2\% was found in $\kappa_{100}$, associated with the variation of the $H$ direction with respect to $\bar{Q}$. The upper bounds of the values of the anisotropy are essentially lower than theoretically predicted variations for the states in which line nodes are running along the FS cylinder only, like a $d$-wave state \cite{makianiz,vekhteraniz}. Therefore our data are in favor of a state with a line node running orthogonal to the FS cylinder, as was proposed recently \cite{hasegawa,balatsky,makianiz}.

Single crystals of Sr$_2$RuO$_4$ were grown by a Ru self flux method in an image furnace \cite{crystals}. The samples for our study were cut from a surface part of a single crystalline rod, to avoid inclusions of the so-called 3-K phase \cite{3K}. The crystals were subjected to annealing to remove residual strain caused by cutting. All the samples were characterized by ac susceptibility and x-ray measurements prior to the thermal conductivity study. The data showed the high crystalline quality by giving the small width of the superconducting transition (10 mK) and the sharp X-ray rocking curve peak limited by the instrumental resolution. For the measurement of $\kappa_{100}$, a sample elongated in the [100] direction with the size of 2$\times$0.5$\times$0.05 mm$^3$ was prepared. For the $\kappa_{001}$ measurement, two samples were prepared with the dimensions 0.5$\times$0.5$\times$2 mm$^3$. We intentionally used samples with the side surfaces running in different directions in the conducting plane to check for the influence of sample demagnetization factors \cite{demagnetization}. We did not find any dependence of the results on the sample orientation within the accuracy of our experiment. 

Usual steady state one- heater-two- thermometers method was used for $\kappa$ measurements \cite{tanatar}. A miniature vacuum cell \cite{cell} with the sample was placed in a double axis rotator in a $^{3}$He refrigerator insert of a superconducting magnet, allowing for a field alignment with an accuracy better than 0.1$^{\circ}$.
The electrical resistivity, $\rho$, and $\kappa$ were measured with the same electrical/thermal contacts. To ensure high resolution of the data it is essential to keep the thermal gradient within a sample low, although this leads to deterioration of the signal to noise ratio. To optimize the accuracy under these countervailing requirements we adopted an intentional overheating equal to 0.005$\times T$, setting the accuracy of our measurement of $\kappa$ at about 0.5\%.
We performed measurements in both fixed $T$- $H$ sweep and fixed $H$-$T$ sweep modes. No difference was noticed between these two sets of data. On a slow $H$ sweep (typically 1 to 10 mT/min) no hysteresis was observed. High precision $H$ sweep measurements take a lot of time and it was difficult to keep the temperature with the long term stability better than 5 mK in varying field and heat load. This limited an accuracy of the $H$-sweep data at about 2\%.

In Fig. 1 we show the temperature dependence of $\kappa_{100}$/$T$ and $\kappa_{001}$/$T$. In the normal state, $\kappa_{100}$ is proportional to $T$, as expected for a normal metal in the $T$ range where the scattering of heat carriers is elastic and determined by impurities \cite{tanatar}. As can be seen from Fig. 1, the anisotropy of thermal conductivity is much lower than the anisotropy of the electrical resistivity. This observation indicates the phononic nature of $\kappa_{001}$ at 1.5 K. In the normal state (in $H$ of 1.5 T parallel to the conducting plane) the $\kappa _{001}$/$T$ curve is linear down to 0.7 K, below some curvature appears. The leading $\kappa \sim T^2$ term in the temperature dependence of the phonon conductivity is naturally ascribed to a transport limited by scattering on conduction electrons \cite{sarma}. 

The electronic contribution to $\kappa_{001}$, $\kappa^{el}_{001}$, in the normal state (in the 1.5 T in-plane field) was estimated in two ways. In the first case we compared magnetoresistance in $\rho_{001}$ and $\kappa_{001}$ \cite{electronfd}. Assuming the same relative change of both $\rho_{001}$ and $\kappa^{el}_{001}$, this estimation gives $\kappa^{el}_{001}$=0.00947 Wm$^{-1}$K$^{-2}$, amounting to 7\% of $\kappa_{001}$ at 1.5 K and 30 \% at 0.3 K. This value almost coincides with $\kappa^{el}_{001}$ determined from the value of $\rho_{001}$ by application of the Wiedemann-Franz ratio, giving 7\% at 1.5 K and 32 \% at 0.3 K. As we see, in the vicinity of $T_{\rm c}$, $\kappa_{001}$ is mainly determined by phonon flow with scattering by conduction electrons. In the superconducting state the density of quasi-particles decreases and $\kappa_{001}$ shows a peak due to the phonon mean free path increase. The magnitude of the peak, is however, 8 times lower than the expectation from the phonon scattering on sample boundaries, showing decisive role of scattering by the electronic carriers even in the superconducting state. Simultaneously the electronic contribution to $\kappa$ is notably suppressed below $T_{\rm c}$ \cite{tanatar}. Therefore, in the superconducting state the phonon contribution is dominating, accounting for 88 \% of $\kappa_{001}$ at 0.3 K \cite{Suzuki}. 

Figure 2 shows the field dependence of both $\kappa _{001}$ and $\kappa _{100}$ at 0.3 K in a magnetic field aligned along the [100] and [110] directions. Despite a clear difference of $H_{\rm c2}$ in the two directions, the slope of $\kappa _{100}$ vs. $H$ at $H_{\rm c2}$ is nearly identical. This observation points to the isotropic gap structure near $H_{\rm c2}$. It is noteworthy that the curves for the two heat flow orientations look mirrorlike. This fact reflects the opposite effect of quasi-particles on the electronic and phononic conductivity, as discussed above. 
Similar curves were obtained with different azimuthal angles $\varphi$ of the field within the plane with a 10$^{\circ}$ step, under exact alignment conditions (with the polar angle {\it $\theta$}=90$^{\circ}$ adjusted for each measurement). The extracted values of the $\kappa _{001}$ at several representative fields, as well as the anisotropy of the upper critical field, are shown in the polar plot of Fig. 3. As can be seen, $H_{\rm c2}$ shows a notable four-fold anisotropy, with the maximum and the minimum in the [110] and [100] directions, respectively. The anisotropy of $H_{\rm c2}$ amounts to 4 \%, in a reasonable agreement with the AC susceptibility \cite{mao} and the in-plane thermal conductivity \cite{tanatar} data. The scatter of $\kappa _{001}$ values in this measurement is about 2\%.

The difference of the magnetic field effect should be the largest for the two principal directions, [100] and [110], within the conducting plane. In view of the tetragonal symmetry of the crystal the superconducting gap can have nodes only in one of these directions. As can be seen from the comparison of the curves in Fig. 2, the difference, if any, is very small. To eliminate the problem of temperature drift during $H$ sweep (which is the essential noise factor) we concentrated on measurements in the $T$ sweep mode. We performed the measurements with fixed $H$ from 0.1 to 1.2 T, with a 0.1 T step, for the [100] and [110] directions. It is essential to cover such a broad field range, since different effects can affect the anisotropy at low and high fields. In Fig. 4 we show the results of these measurements at some representative fields, presented as deviations from the average $\kappa$ value. We find that the deviations are random and less than 0.5 \%.

In order to understand this result, first we recall the anisotropy in the quasiparticle density generated with field, predicted for a $d$-wave state by Vekhter {\it et al.} \cite{Vekhterspecheat}, with the difference for [100] and [110] directions amounting up to 30 \%. To our knowledge there is no experimental observation of this effect. For the cuprates the limitation in experimentally accessible values of the reduced magnetic field, $H/H_{\rm c2}$ $\ll$ 1 is one of the possible reasons. For Sr$_2$RuO$_4$, in which the whole field range is accessible, the anisotropy of the $H_{\rm c2}$ itself is making a dominant contribution in the very vicinity of $H_{\rm c2}$. At 1.2 T, near (but not very close) to $H_{\rm c2}$, the anisotropy is already very small. Since the interplane heat flow is influenced only by the quasiparticle density, the latter remains constant within the accuracy of our experiment. Thus such anisotropy effect is not present here. 
Another contribution is the anisotropy in the scattering of quasi-particles due to the Andreev reflection on the superfluid current around the vortex core. This effect would be important for electronic contribution at low fields and is responsible for the weak-field anisotropy observed in the cuprates \cite{yuaniz,4foldcuprates}. Again, there is no evidence for this effect in Sr$_2$RuO$_4$.

In this Letter, we evaluated the anisotropy by covering the whole range of the field. The lack of the anisotropy both at low and high fields makes it reasonable to conclude that both of the abovementioned mechanisms are not operating here. 
The isotropic quasi-particle contribution can come either from a line node in the superconducting gap running orthogonal to the FS cylinder, or from a sheet of the FS whose superconducting gap is easily and isotropically destroyed by the in-plane field. The former situation is quite analogous to the case of UPt$_3$, in which the presence of the horizontal node is a likely reason for the lack of anisotropy \cite{SuderowUPtaniz}. The latter is a natural extension of the orbital - dependent superconductivity (ODS) model \cite{sigrist2}.
From the anisotropy experiment alone it is not possible to make the preference for either the ODS or the "horizontal" node model. However, for a full gap state it is difficult to explain the power-law dependence of the quasiparticle density at low temperatures \cite{nqr,penetration}. Therefore, an ODS model with full gaps on all three FS sheets does not give an appropriate description of the complete set of experiments. Measurements at lower temperatures may clarify this issue.

In conclusion, we have found no notable anisotropy of the thermal conductivity in Sr$_2$RuO$_4$ with high experimental resolution. In conjunction with the existence of low temperature quasi-particle excitations, indicative of nodes in the gap, our results strongly suggest the presence of the line node running orthogonal to the Fermi surface cylinder.

The authors acknowledge A. E. Kovalev and V. A. Bondarenko for the help in experiment design, and D. Agterberg, A. G. Lebed, K. Maki and M.Sigrist for valuable discussions. 

{\bf Note added}. In the following Letter, K.Izawa {\it et al.} \cite{Izawa} make an important extension of our previous study of anisotropy in $\kappa_{100}$ \cite{tanatar}.

\begin{figure} 
\begin{center}
	\epsfxsize=6cm
	\epsfbox{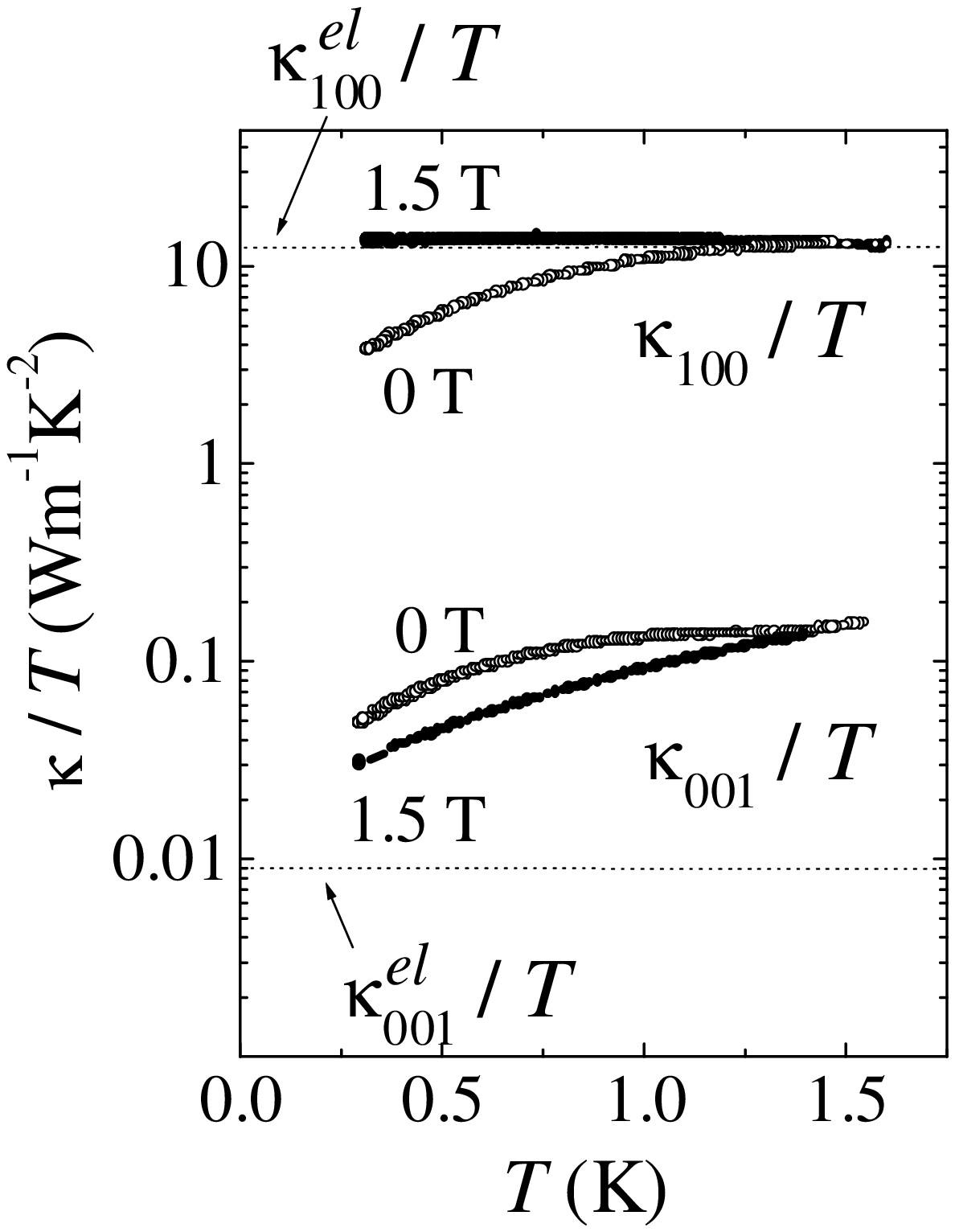}
\end{center}
\caption{Temperature dependence of the in-plane, along [100], and the interplane, along [001], thermal conductivities of Sr$_{2}$RuO$_4$ in zero magnetic field and in the normal state. Note the logarithmic scale for $\kappa/T$. The dotted lines represent electronic thermal conductivity for respective configurations estimated from the Wiedemann-Franz law.} 
\end{figure}

\begin{figure} 
\begin{center}
	\epsfxsize=6cm
	\epsfbox{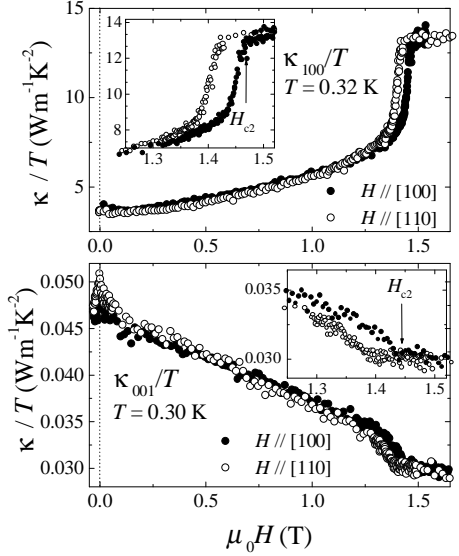}
\end{center}
\caption{ Field dependence of thermal conductivities $\kappa_{001}$ and $\kappa_{100}$ at 0.3 K for the magnetic field along two principal directions, [100] and [110], in the conducting plane. Insets show an expanded view of a region near $H_{rm c2}$.}
\end{figure}

\begin{figure} 
\begin{center}
	\epsfxsize=7cm
	\epsfbox{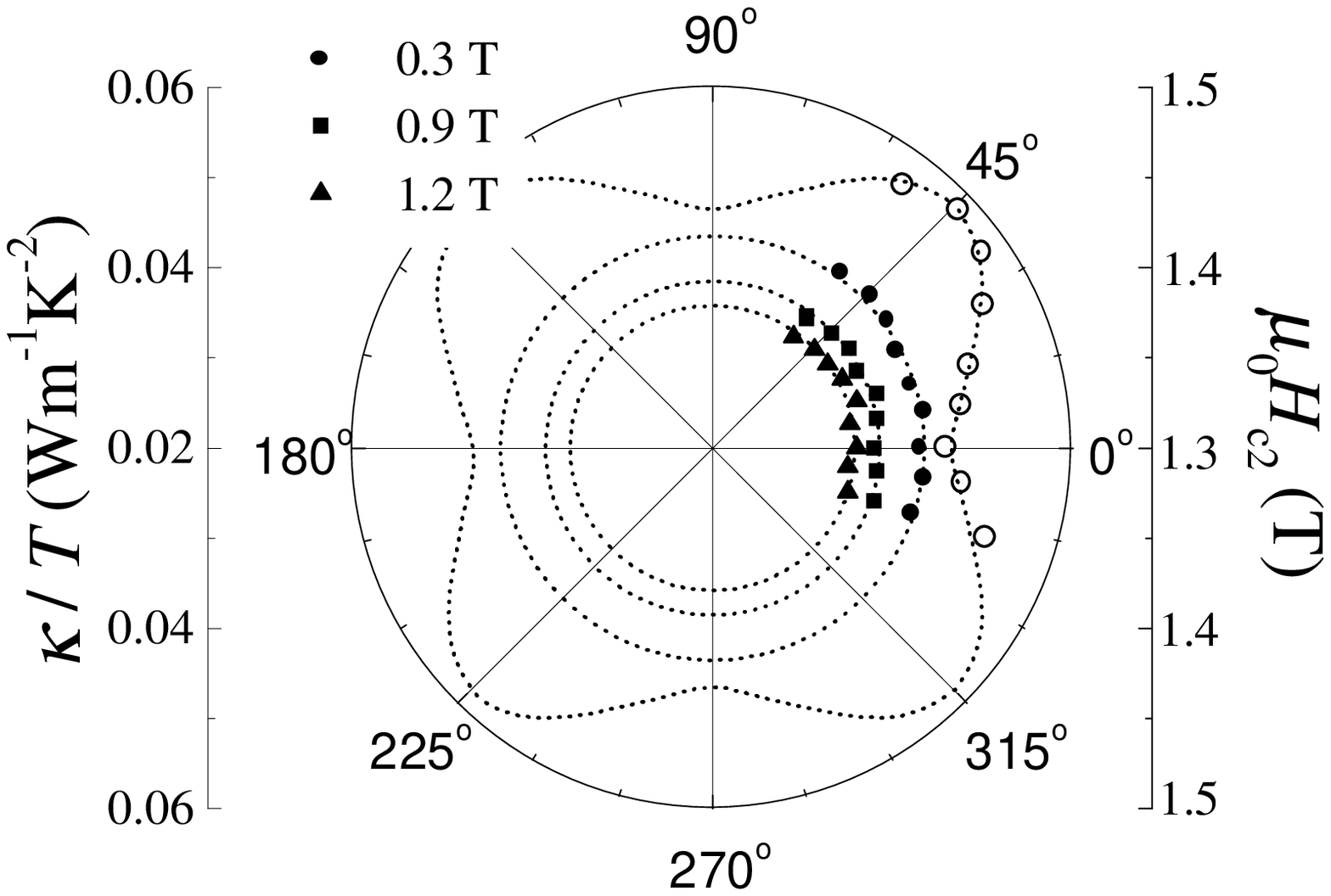}
\end{center}
\caption{ Polar plot of $\kappa_{001}$ at 0.3 K for several values of the field (closed symbols), overlaying the polar plot of the upper critical field (open symbols). Note the expanded scales. } 
\end{figure}

\begin{figure} 
\begin{center}
	\epsfxsize=7cm
	\epsfbox{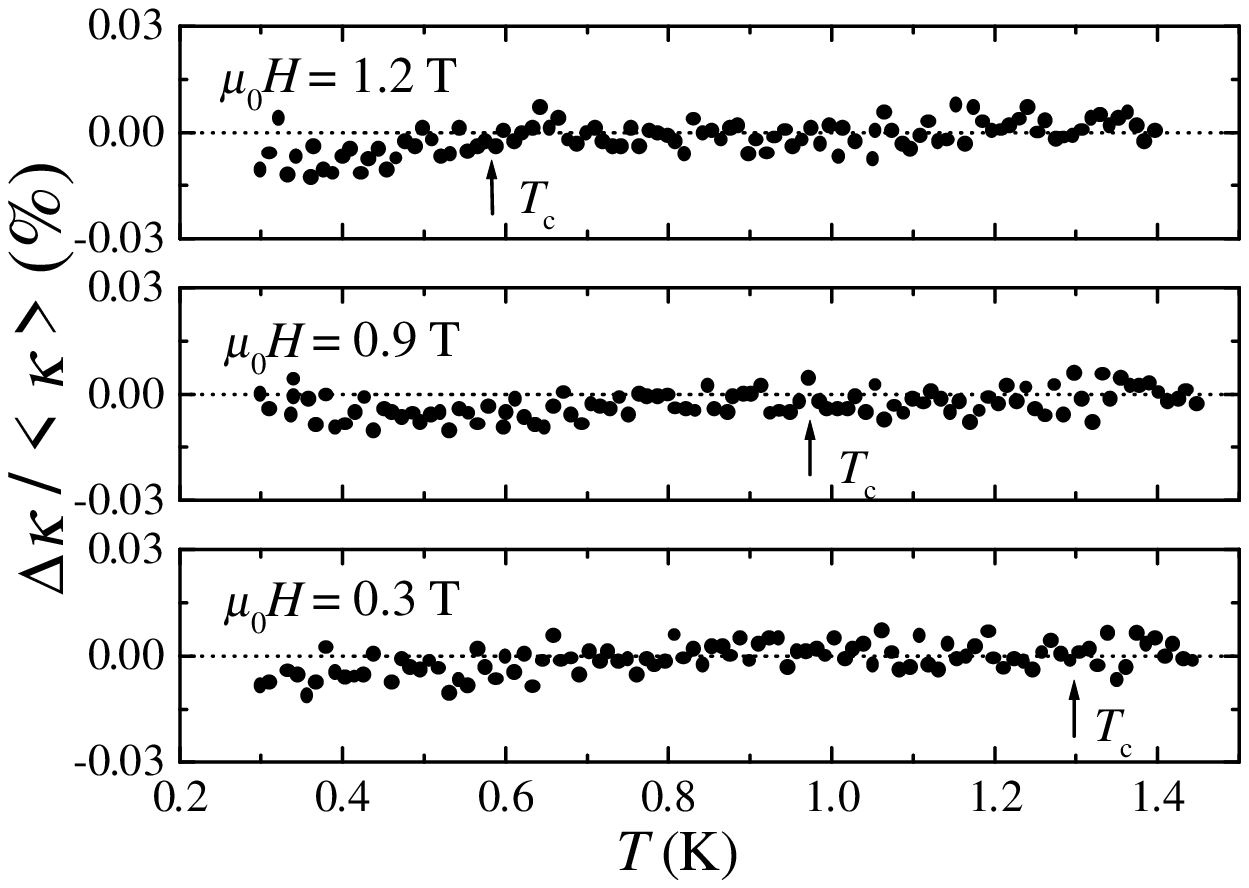}
\end{center}
\caption{ Temperature dependence of the relative difference between the interplane thermal conductivities for [100] and [110] field directions for some representative fields. $\Delta \kappa /\langle \kappa \rangle $$\equiv$[$\kappa_{001}^{H \| [110]}$($T$)-$\kappa_{001}^{H \| [100]}$($T$)]/[$\kappa_{001}^{H \| [110]}$($T$) + $\kappa_{001}^{H \| [100]}$($T$)]. Arrows show superconducting $T_{\rm c}$ for each value of the magnetic field. }
\end{figure}

\end{document}